# Infrared single-pixel imaging utilising microscanning

**Ming-Jie Sun**[1,2,*], **Matthew P. Edgar**[2], **David B. Phillips**[2], **Graham M. Gibson**[2], **and Miles J. Padgett**[2]

1 Department of Opto-electronic Engineering, Beihang University, Beijing, 100191, China
2 SUPA, School of Physics and Astronomy, University of Glasgow, Glasgow, G12 8QQ, UK
*sunmingjiecn@163.com

**ABSTRACT**

Since the invention of digital cameras there has been a concerted drive towards detector arrays with higher spatial resolution. Microscanning is a technique that provides a final higher resolution image by combining multiple images of a lower resolution. Each of these low resolution images is subject to a sub-pixel sized lateral displacement. In this work we apply the microscanning approach to an infrared single-pixel camera. For the same final resolution and measurement resource, we show that microscanning improves the signal-to-noise ratio (SNR) of reconstructed images by $\sim 50\%$. In addition, this strategy also provides access to a stream of low-resolution 'preview' images throughout each high-resolution acquisition. Our work demonstrates an additional degree of flexibility in the trade-off between SNR and spatial resolution in single-pixel imaging techniques.

**Introduction**

Conventional digital cameras use a lens system to form an image of a scene onto a detector array. The spatial resolution of the recorded image can be limited either by the point spread function of the optical system, or by the pitch of the pixels in the detector array. If the resolution is limited by the pixel pitch, the most common method to improve this is to increase the number of pixels per unit area by reducing their physical size. However, apart from the technological challenges associated with this approach, smaller pixels detect less light, which degrades image quality.[1] An alternative approach to increase the pixel resolution is microscanning.[2–6] In this approach, multiple images of the same scene are recorded, while the detector array is displaced by sub-pixel sized translations between images. Data from several images is then combined to reconstruct a composite image with a spatial resolution exceeding that of the detector array.

While there has been a global drive to increase the number of pixels in camera sensors, there has also been significant developments in camera technology that records images using just a single-pixel detector.[7–9] These so called 'single-pixel cameras' make use of spatial light modulators (SLMs, a technology made affordable by huge consumer demand) to sequentially apply a series of binary masks to the image. The single-pixel detector records the total intensity transmitted through each mask. Knowledge of the transmitted intensities and the corresponding masking patterns enables reconstruction of the image. This approach is of particular interest when the operational wavelength renders the development of detector arrays prohibitively expensive or impossible.[10, 11]

Within the single-pixel camera approach, it is the spatial resolution of the masking patterns that is equivalent to the pixel resolution. The number of masks required to reconstruct a fully sampled image increases with the square of the resolution (i.e. in proportion to the total number of pixels in the reconstructed image). However, in single-pixel imaging systems there is also a further restriction: the signal-to-noise ratio (SNR) of the reconstructed image decreases as the resolution is increased. This is because typically masks are 50 % transmissive, and consequently the fluctuations in the transmitted intensity from one mask to another occur about a large DC offset. As the resolution of the mask is increased, the relative size of the fluctuations, which contain the information, is reduced (roughly in proportion to the square root of the number of pixels in the mask). It is the size of these fluctuations compared to the detector noise that sets a limit to the useful mask resolution, i.e. in a single-pixel camera a higher resolution mask set gives a higher resolution image, but one with a lower SNR. The trade-off between the image resolution and its SNR restricts further applications of single-pixel camera technology. To improve this situation, schemes such as differential ghost imaging,[12–16] have been developed to reduce the noise without jeopardizing resolution.

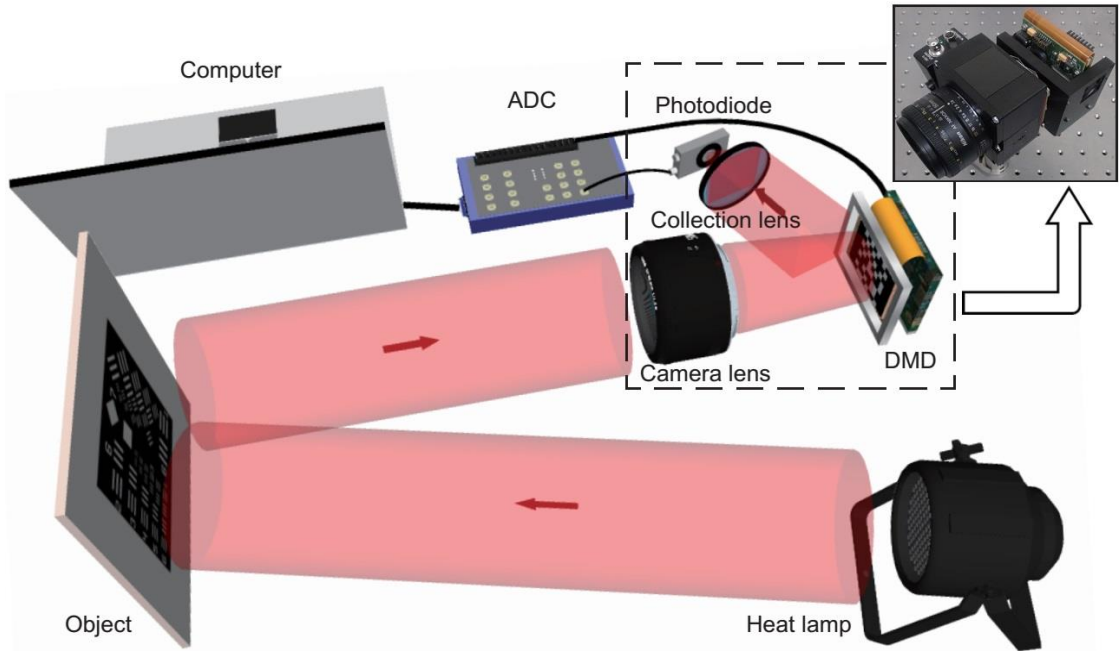

**Figure 1. Experimental set-up.** A heat reflector lamp illuminates the object, which is a 100mm×100mm grey-scale picture located at a distance of ∼ 0.5 m from the imaging system. A 50 mm camera lens collects the reflected near-infrared light and images the object onto a high-speed digital micro-mirror device (DMD). The DMD is placed at the image plane and applies rapidly changing binary masks to the transmitted image. An InGaAs detector measures the total intensity transmitted through the masks. An analogue-to-digital converter (ADC), triggered by the synchronisation TTL signals from the DMD, acquires and transfers the light intensities data to a computer for image reconstruction.

In this work, we show that it is possible to enhance the pixel resolution of an infrared single-pixel camera, while maintaining its SNR, by adopting a microscanning approach. We obtain multiple low-resolution images, each laterally shifted by sub-pixel steps. These images are then co-registered on a higher resolution grid to give a single high-resolution image. We demonstrate that for the same measurement resource (i.e. same number of masks and acquisition time), this methodology results in a reduced noise level in the final high-resolution reconstructed image for only a slight reduction in resolution. In addition, our method simultaneously delivers a sequence of low-resolution 'preview' images *during* the high-resolution image acquisition. We note that the concept of microscanning in the context of single-pixel imaging was recently considered theoretically in Ref. 17. Our microscanning technique is applicable to all single-pixel imaging systems, both those based on projected light fields and those based on image masking, the latter of which is demonstrated here. More importantly, microscanning in the single-pixel camera context requires no extra hardware and only a trivial increment in algorithm complexity. Therefore, our method can be deployed as a complement to existing single-pixel camera systems, including those utilising compressive sensing.[7,18,19]

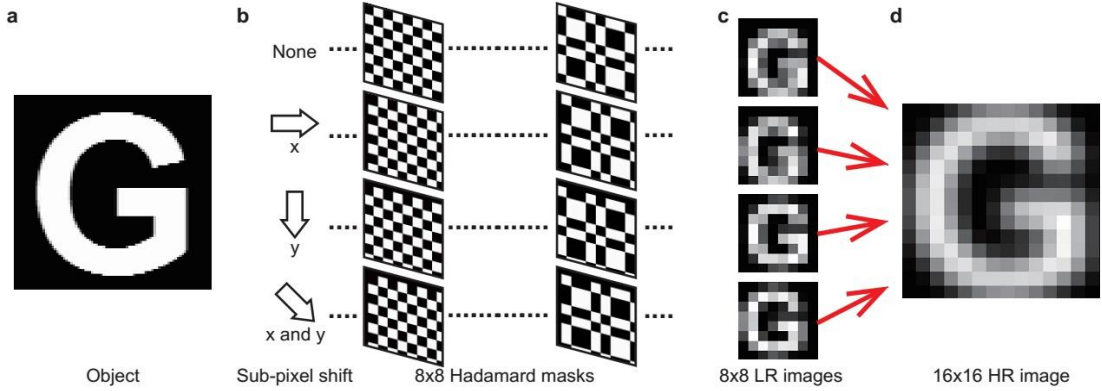

**Figure 2. The full-pixel microscanning method.** (a) Binary object to be imaged. (b) A set of low-resolution (LR) Hadamard masks is displayed four times, with a lateral shift in *x* and/or *y* of half a pixel width applied between each mask set. (c) This operation yields four low-resolution 'preview' images, each of which contains different spatial information. (d) This allows a high-resolution (HR) image to be reconstructed by co-registering each image in its laterally shifted location on a higher resolution grid.

**Experimental setup**

Figure 1 illustrates our infrared single-pixel camera set-up, which is a modified and compacted version of the system in our previous work.[11] We test our system by imaging a grey-scale picture, which is located ~0.5m from the single-pixel camera and illuminated with a heat reflector lamp. The camera lens images the scene onto a high-speed digital micromirror device (DMD) which is used to sequentially mask the image of the scene with a preloaded sequence of binary masks. The total intensity of light transmitted through each mask is detected by a InGaAs detector. A high dynamic range analogue-to-digital converter (ADC), synchronised with the DMD, acquires and transfers the intensity data to a computer for image reconstruction. A photograph of our integrated single-pixel camera prototype is shown in Fig. 1, and a detailed specification of the system is provided in the Methods.

As in our previous work, we make use of Hadamard matrices to form our DMD masking patterns.[20,21] This is a convenient basis, as the rows (or columns) of the Hadamard matrices form a complete orthogonal set, enabling efficient sampling of the image at a well defined resolution using a given set of patterns.[11,22] The elements of the Hadamard matrices take values of '1' or '-1', and each row is reformatted into a 2D grid and displayed as a 2D binary mask on the DMD, where '1' and '-1' denote micromirrors 'On' and 'Off' respectively. Therefore, the intensity signal $S_p$ associated with a particular masking pattern $M_p$ is given by:

$$S_p = \sum_i \sum_j (M_{p,ij} \cdot D_{ij}), \quad (1)$$

where *i* and *j* index the *x* and *y* coordinates of the binary mask respectively, and *D* is the intensity distribution on the image plane that we wish to reconstruct, also discretized by the mask pixels. After performing *n* measurements, the image *I* can be reconstructed as

$$I = \sum_{p=1}^{n} (M_p \cdot S_p). \quad (2)$$

In order to reduce sources of noise such as fluctuations in ambient light levels, we obtain differential signals by displaying each Hadamard mask immediately followed by its inverse (where the micromirror status 'On' and 'Off' are reversed), and taking the difference in the measured intensities.[12,13,15]

In our experiments, we utilise the square central region of our DMD, which consists of 768×768 micromirrors. The DMD micromirrors are grouped into 'super-pixels' to display the reformatted Hadamard masks. For example, if displaying an 8×8 pixel Hadamard mask, each Hadamard pixel therefore comprises of 96×96 micromirrors. In order to apply microscanning to single-pixel imaging, it is the position of the masks on the DMD that must be laterally displaced by sub-pixel translations. In our results, we compare two different microscanning approaches, based on the percentage of active micromirrors used within each pixel, which are detailed below. We contrast both of these microscanning methods with the standard technique of increasing the image resolution by simply increasing the resolution of the Hadamard mask set, which we refer to here as 'normal' high-resolution (NHR) imaging.

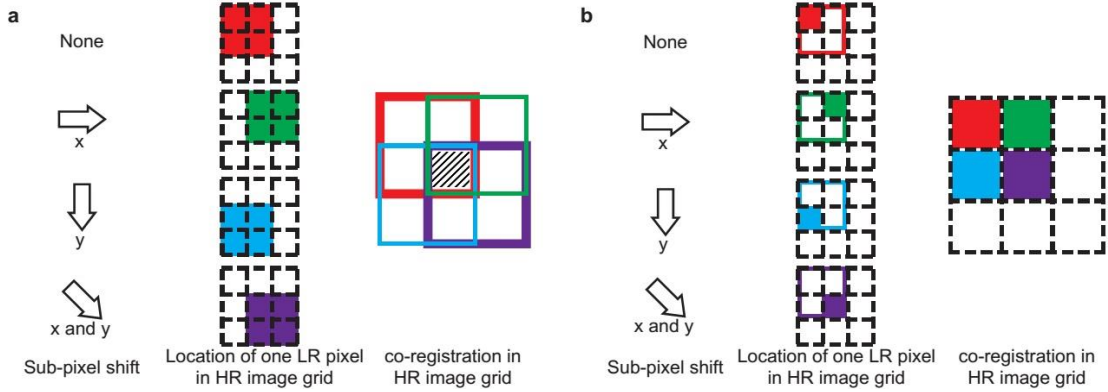

**Figure 3. Schematics of our two microscanning strategies.** (a) Full pixel microscanning. (b) Quarter pixel microscanning.

## Results

**The full-pixel microscanning (FPM) method** uses lower resolution masks which utilise all of micromirrors within each Hadamard pixel. Figure 2 illustrates the reconstruction of a 16×16 pixel image from four 8×8 pixel images of the equivalent field-of-view. A set of 8×8 pixel Hadamard masks (each pixel comprising of 96×96 DMD micromirrors) is displayed four times, with a lateral shift in $x$ and/or $y$ of half a pixel width (i.e. 48 DMD micromirrors) applied between each set (Fig. 2(b)). This operation yields four 8×8 pixel images (Fig. 2(c)), using the same number of masks as required to obtain a 16×16 pixel NHR image. Each of these 8×8 pixel images contains different spatial information. A 16×16 pixel image is then reconstructed by co-registering each 8×8 pixel image in its laterally shifted location on a 16×16 grid, and averaging the overlapping pixel values for each high-resolution pixel (Fig. 2(d)). Therefore, as shown in Fig. 3(a), the value of the hatched central pixel on the high-resolution grid is computed as:

$$I_{central} = \frac{1}{4}(I_{red} + I_{green} + I_{blue} + I_{purple}),  \qquad (3)$$

where $I_{red}$, $I_{green}$, $I_{blue}$, and $I_{purple}$, represent the measured intensities in the low-resolution pixels as shown. Therefore, each pixel in the high-resolution reconstruction is computed from a unique set of low-resolution pixels. In the absence of noise, the reconstructed result using FPM is mathematically equivalent to the convolution of the NHR image (obtained in the standard way using a sequence of 16×16 Hadamard masks in this example) with a kernel:

$$\mathcal{K} = \frac{1}{16}\begin{bmatrix} 1 & 2 & 1 \\ 2 & 4 & 2 \\ 1 & 2 & 1 \end{bmatrix}. \qquad (4)$$

This convolution causes a modest reduction in the contrast of the high spatial frequencies in the reconstructed FPM image (see Fig. 4). However, FPM does offers a significantly improved SNR compared to the NHR method, because the SNR of a single-pixel camera image decreases roughly in proportion to the square root of the number of mask pixels. Therefore a high-resolution image reconstructed from four lower-resolution images inherits the higher SNR of the lower-resolution images.

We also note that it is possible to use a matrix inversion to recover the NHR image from the four laterally shifted low resolution images. The matrix inversion method is equivalent to solving a set of simultaneous equations describing the intensity of each high-resolution pixel in terms of the low resolution measurements. However, matrix inversion requires the use of appropriate boundary conditions at the edges of the image, and we found this method to be highly unstable with respect to small levels of noise in the measurements.

**The quarter-pixel microscanning (QPM) method** again allows reconstruction of a higher resolution image from four half-pixel laterally shifted images. However in this case, each lower resolution image is recorded with only one quarter of the micromirrors active within each pixel, as shown in Fig. 3(b). The four quadrants of the pixel are sequentially utilised, and each quadrant corresponds to a half-pixel shift in *x* and/or *y*. This ensures that each low-resolution image samples a non-overlapping region of the scene. Therefore, the four low-resolution images obtained using QPM can simply be co-registered accordingly on a high-resolution image grid, with no need for averaging. Consequently, the resolution of the QPM image is identical to that of the NHR image in the absence of noise. As with FPM, QPM also delivers low-resolution images within each high-resolution image acquisition. However, as only a quarter of the micromirrors within each pixel are active during the acquisition, QPM suffers from a reduced SNR compared to FPM and NHR.

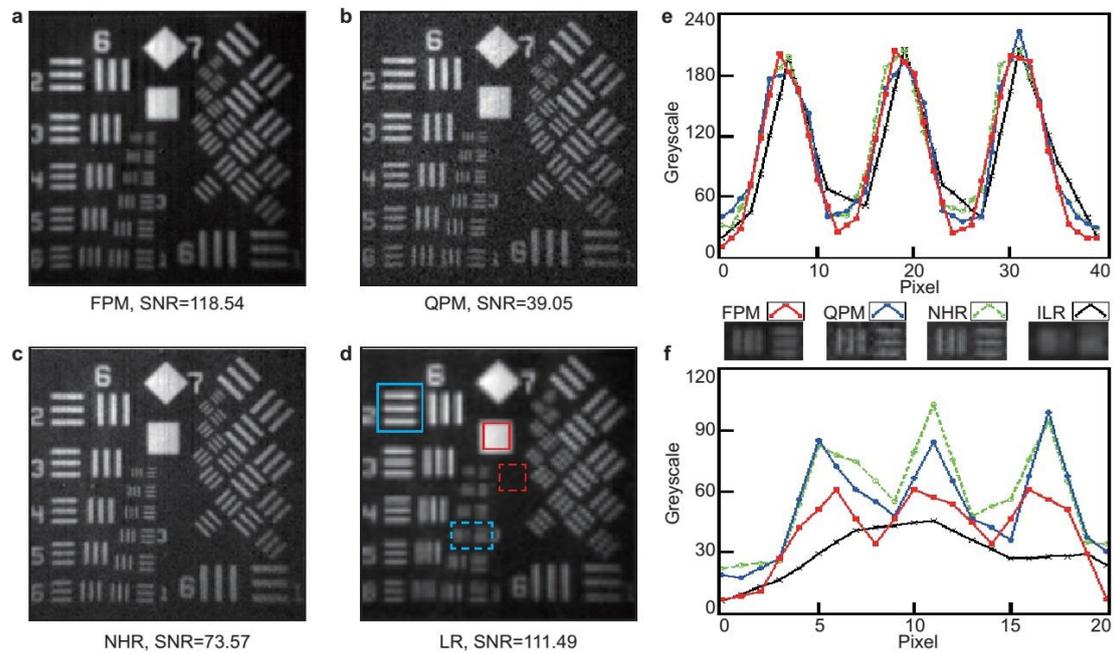

**Figure 4. Experimental images of a modified USAF resolution test chart.** Images reconstructed using (a) full-pixel microscanning (FPM), (b) quarter-pixel microscanning (QPM), (c) normal high-resolution (NHR) method, and (d) low-resolution (LR) method. In (d) a $64 \times 64$ pixel image has been interpolated upto a $128 \times 128$ pixel image for comparison. The experimentally measured SNR is quoted beneath each image. (e) Line profiles through the regions highlighted by the solid blue box. (f) Line profiles through the regions highlighted by the dashed blue box. Insets show magnified views of the dashed blue box regions for each imaging method.

Figure 4 shows a comparison of images recorded using the different approaches. To test the performance of each method, we image a modified US Air Force (USAF) resolution test chart. Figures 4(a-c) show $128\times 128$ pixel images reconstructed using FPM (a), QPM (b), and NHR (c). Figure 4(d) shows an example of a $64 \times 64$ pixel image, linear interpolated into $128\times 128$ resolution for comparison. For each image we calculate the SNR using:[23, 24]

$$\text{SNR} = (\langle I_f \rangle - \langle I_b \rangle)/((\sigma_f + \sigma_b)/2), \qquad (5)$$

where $\langle I_f \rangle$ is the average intensity of the feature (here calculated from the data within the white block, highlighted by a solid red square in Fig. 4(d)), $\langle I_b \rangle$ is the average intensity of the background (here calculated from the data highlighted by the dashed red square in Fig. 4(d)), and $\sigma f$ and $\sigma b$ are the standard deviations of the intensities in the feature and the background respectively.

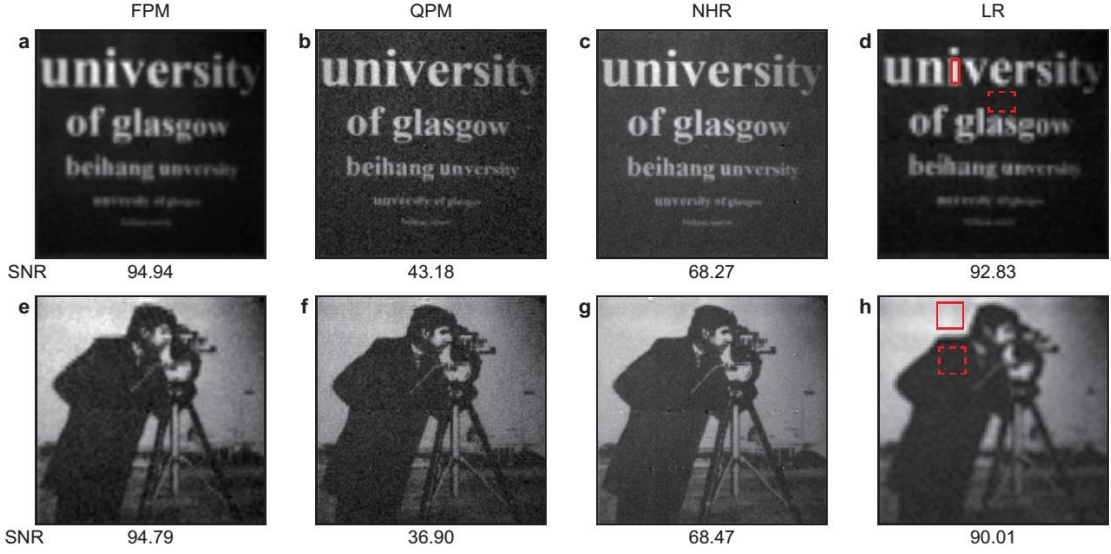

**Figure 5. Experimental images of two different objects.** (a-d) Images of letters with different sizes obtained using different methods. (e-h) Images of 'Cameraman' obtained using different methods. The experimentally measured SNR quoted beneath each image is computed using the highlighted features (solid red box) and background (dashed red box).

The SNR of the image reconstructed using FPM is significantly higher than that of the image obtained using NHR. Figures 4(e) and (f) show line profiles through two different regions of the USAF test chart containing different spatial frequencies. In Fig. 4(e), the features (highlighted by the blue solid box in Fig. 4(d)) are well resolved using all four imaging methods. In Fig. 4(f), the vertical line profiles are taken through the features highlighted by the blue dashed box in Fig. 4(d). Magnified views of these regions are shown as figure insets to Fig. 4(f). In this case, the features are now only resolved in the images obtained using NHR, QPM and FPM. The effect of the convolution is evident in the FPM linescan, which displays a modest reduction in contrast compared to NHR and QPM. Figure 5 demonstrates that the trends in image SNR discussed above are also reproducable when imaging different objects.

In low light levels, the advantage of improved SNR using FPM becomes more significant. For example, in Figs. 6(a-c) the noise is so severe when using the NHR method that parts of the image are unidentifiable. However, under identical illumination conditions, FPM enables a dramatic improvement in image quality, as shown in Figs. 6(g-i). As discussed above, in the absence of noise FMP is mathematically equivalent to NHR convolved with the smoothing Kernel $\mathcal{K}$. However even when the NHR images are themselves smoothed with Kernel $K$, the resulting image quality is still far lower than the images obtained using FMP.

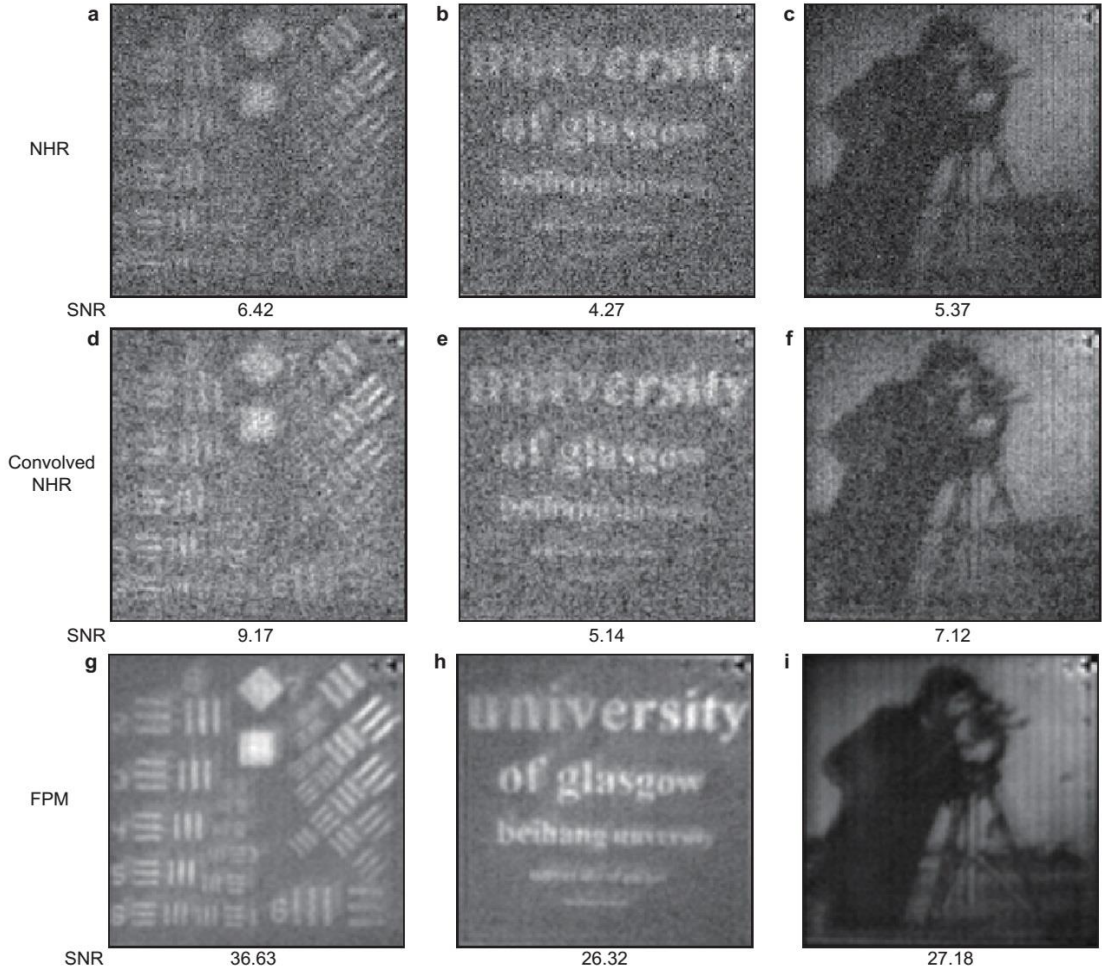

**Figure 6. Comparison of images obtained in a high noise situation.** (a-c) Images obtained using NHR. (d-f) Images shown in (a-c) are here convolved with kernel $\mathcal{K}$ (Eq. (4)). (g-i) Images obtained under identical conditions using FPM. The measured SNR (using the same features and background as before) is quoted beneath each image, demonstrating the superior performance of FPM in this situation.

While FMP offers advantages at low light levels, the QPM may find more use operating in high light conditions. For example, in our experiments the photodiode gain is set in order to maximise the dynamic range of the ADC. As the QPM method reduces the amount of light arriving at the photodiode, the gain can be increased in this situation (assuming it has not already been maximised), to improve the QPM image quality. This is demonstrated in Fig. 7 where (a-c) show images obtained using QPM with a 40 dB gain (equivalent to the gains used in all images shown in Figs. 4 and 5), while Figs. 7(d-f) show the improvement in SNR when using a gain of 50 dB with the QPM method.

**Discussion**

We have demonstrated two image reconstruction strategies based on a microscanning approach applied to an infrared single-pixel camera. Both of these strategies deliver a sequence of low-resolution 'preview' images throughout the high-resolution image acquisition. For example we obtain $128 \times 128$ pixel images at a rate of $\sim 0.5$ Hz (using 32,768 Hadamard masks displayed by the DMD at 20 kHz). During the high-resolution acquisition our method delivers $64 \times 64$ pixel 'preview' images at a rate of ~2 Hz. Compared to conventional sampling, the full-pixel microscanning strategy improves the SNR of images, at the expense of a modest reduction in the contrast of high spatial frequencies. The quarter-pixel microscanning strategy suffers from reduced SNR, which under certain situations can be improved by optimizing the photodiode gain. Our work demonstrates an additional degree of flexibility in the trade-off between SNR and spatial resolution in single-pixel imaging techniques, which can be optimized as the case demands.

Our work is applicable to all single-pixel imaging systems, both those based on structured illumination and those based on masked detection, the latter of which is demonstrated here. Importantly, our approach requires no additional hardware and can be utilised as a complement to existing single-pixel imaging schemes, such as compressive sensing. It is also worth noting that the microscanning technique is often utilised in infrared imaging because the larger pixel pitch (relative to visible wavelength detector arrays) leads to more severe pixel aliasing. Considering the cost of infrared detector arrays, our work demonstrates the potential for a cost-effective and flexible solution to achieve high-resolution imaging in the infrared.

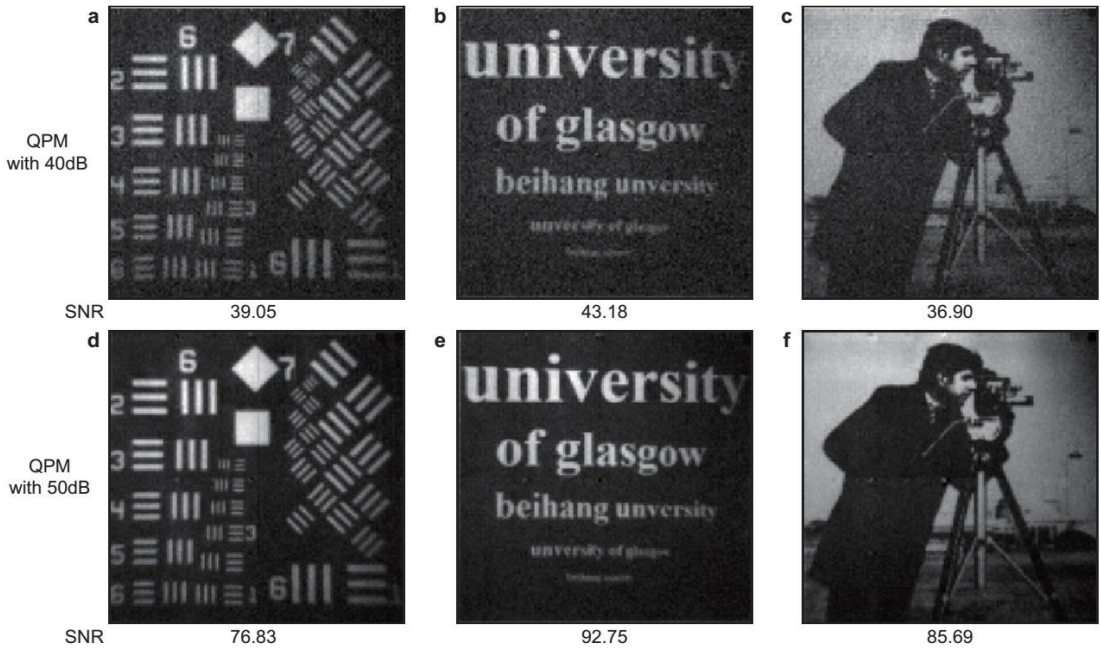

**Figure 7. Comparison of images obtained using QPM with different gains.** (a-c) 40 dB gain. (d-f) 50 dB gain. Experimentally measured SNR (using the same features and background as before) is quoted beneath each image.

**Methods**

In our experiments, a heat reflector lamp (Philips PAR38 IR 175C) illuminates the object, which is a 100 mm × 100 mm grey-scale picture located at a distance of ~0.5 m from the imaging system. A 50 mm camera lens (Nikon AF Nikkor, f/1.8D) collects the reflected near-infrared light and images the object onto a DMD (Texas Instruments Discovery 4100, 1024 × 768, operating at 20kHz) functioning as a SLM. An InGaAs detector (Thorlabs PDA20CS InGaAs, 800 nm to 1800 nm, 0 dB-70 dB gain) measures the total intensity transmitted through the masks. An ADC (National Instruments DAQ USB-6221 BNC, sampling at 250 kS/s), triggered by the synchronisation TTL signals from the DMD, acquires and transfers the light intensities data to a computer for image reconstruction.

**Acknowledgements**

MS and DBP thank Dr. Jonathan Taylor for helpful discussion. MS acknowledges the support from National Natural Science Foundation of China (Grant No. 61307021) and China Scholarship Council (Grant No. 201306025016). MJP acknowledges financial support from the Wolfson foundation and the Royal Society.

**Author contributions statement**

MS and MJP conceived the concept of the experiment. MS, MPE and GMG conducted the experiments. MS, DBP and MJP designed the reconstruction algorithm and analysed the results. All authors contributed to writing the manuscript.

**Additional information**

**Competing financial interests:** The authors declare no competing financial interests.